# Enhanced Tunnel magnetoresistance in Fe/Mg$_4$Al-O$_x$/Fe(001) Magnetic Tunnel Junctions

Thomas Scheike, Zhenchao Wen, Hiroaki Sukegawa,[a] and Seiji Mitani

*National Institute for Materials Science, 1-2-1 Sengen, Tsukuba 305-0047, Japan*

## Abstract

Spinel MgAl$_2$O$_4$ and family oxides are emerging barrier materials useful for magnetic tunnel junctions (MTJs). We report large tunnel magnetoresistance (TMR) ratios up to 429% at room temperature (RT) and 1,034% at 10 K in a Fe/MgAl$_2$O$_4$/Fe(001)-based MTJ prepared using electron-beam evaporation of Mg$_4$Al-O$_x$. Resistance oscillations with a MTJ barrier thickness of 0.3-nm were significantly enhanced compared to those of a Fe/MgO/Fe(001) MTJ, resulting in a large TMR oscillation peak-to-valley difference of 125% at RT. The differential conductance spectra were symmetric with bias polarity, and the spectrum in the parallel magnetization state at low temperature demonstrate significant peaks within broad local minima at |0.2–0.6| V, indicating improved barrier interfaces by the Mg$_4$Al-O$_x$ barrier. This study demonstrates that TMR ratios in Fe(001)-MTJs can still be improved.

[a] email: sukegawa.hiroaki@nims.go.jp

A magnetic tunnel junction (MTJ) is the basic building block for various spintronic applications,[1–4] such as magnetoresistive random access memories, read-heads of hard disk drives, and high-sensitivity magnetoresistive sensors. They consist of at least two magnetic layers separated by an insulator (barrier). Because of tunnel magnetoresistance (TMR) effects, the relative orientation of the magnetization of their magnetic layers results in high and low resistance states when the magnetization is parallel (P) or antiparallel (AP). A large TMR ratio is desired for novel spintronic applications that require a fast and stable operation.[5–8] Fe/MgO/Fe(001) MTJ is a simple and standard MTJ structure with large TMR ratios of 180%–220% at room temperature (RT),[9–12] which originate from a $\Delta_1$-band preferential coherent tunneling mechanism.[13,14] However, even at low temperatures (LT), the experimental TMR ratio in Fe/MgO/Fe(001) (250%–370%) was much smaller than the predicted values in first-principles calculations (>1,000%).

Recently, some authors demonstrated a TMR ratio of 417% at RT (914% at 5 K) in the epitaxial Fe/MgO/Fe(001),[15] which is almost twice (three times) as large at RT (LT) as the values in previous reports on Fe/MgO/Fe. The significantly increased values are the result of high-quality epitaxial growth and chemically sharp interfaces of the multilayer structures, which was confirmed by well-developed deep minima in the differential conductance spectra. Especially, MgO barrier interface modifications combining electron-beam (EB) evaporation and post-oxidation techniques were critical for improving interface crystallinity. However, we found that many misfit dislocations were still introduced at the MgO bottom and top interfaces because of a large lattice mismatch between MgO and Fe (3.8% for bulk). This means that there is less room to improve the multilayer quality near the barrier when using a MgO barrier for Fe electrodes. A significant lattice distortion is induced in the vicinity of misfit dislocations, which can lead to undesirable electron scattering; therefore, improving the current TMR ratio using only Fe and MgO proves difficult.

MgAl$_2$O$_4$ (MAO) spinel is a good barrier material for reducing such interfacial dislocations because of the perfect lattice-matching between MAO and Fe.[16–18] MAO-based MTJs improve TMR ratios by introducing cation-disordering into the spinel structure (cation-disordered MgAl$_2$O$_4$),[19–21] especially at high bias voltages.[17,22] Importantly, the atomic ratio of Mg and Al in MAO is tunable for a wide composition range (i.e., Mg-Al-O) while maintaining its cubic structure,[3] resulting in lattice-matched interfaces with various ferromagnetic layers.[16–18,23] Recently, a relatively large TMR ratio of up to 240% was observed in lattice-matched MTJs with a textured CoFeB/MAO/CoFeB structure using RF sputtering of sintered MAO with an Mg-rich composition (Mg/Al atomic ratio = 2).[24] The Mg-rich MAO barrier has the advantage of easily achieving good lattice-matching using the fabrication methods used for a MgO barrier.

In this letter, we report an improved TMR ratio of up to 429% at RT by developing a Fe/Mg-rich MAO/Fe(001) epitaxial MTJ structure using EB-evaporation and magnetron sputtering for the Mg$_4$Al-O$_x$ barrier and Fe electrode, respectively. At LT, the TMR ratio exceeded a thousand percent (up to 1,034% at 10 K), reaching a value predicted by theoretical calculations of Fe/MgO/Fe(001). The differential conductance spectrum of the prepared MAO-based MTJs in the P state shows distinct local structures at LT, which are more significant than those of the recently reported Fe/MgO/Fe MTJ. Furthermore, the MAO-based MTJ shows a significant oscillatory TMR ratio with barrier thickness as the same oscillation period as Fe/MgO/Fe (~0.3 nm) and a large peak-to-valley difference of 125% at RT. These features can be attributed to improved barrier interface structures caused by the lattice-tuning effect of the MAO barrier.

MTJ structures were fabricated at RT using magnetron sputtering and EB-evaporation at a base pressure of 4×10$^{-7}$ Pa on single crystal MgO(001) substrates. The typical stack structure is MgO substrate/Cr (60)/Fe (50)/Mg (0.5)/wedge-shaped MAO ($d_{bar}$ = 1.0–3.0)/natural oxidation/Fe (5)/Ir$_{22}$Mn$_{78}$ (IrMn) (10)/Ru (20) (thickness in nm) (Fig. 1). The MAO barrier was deposited using EB-evaporation of a sintered MAO block with a nominal Mg/Al atomic ratio = 4 ($Mg_4Al-O_x$). Back-pressure and deposition rates were ~1 × $10^{-5}$ Pa and 8 × $10^{-3}$ nm/s, respectively. The wedge-shaped MAO layer was prepared using a linear motion shutter. Cr-buffer and bottom- (top-) Fe were annealed at 600°C and 300°C (400°C), respectively, as described in Ref. [15]. The barrier was post-annealed at 250°C and after cooling to RT, naturally, in-situ oxidized using pure $O_2$ gas (99.999%, ~5 Pa) for 300s. After the deposition, the unpatterned films were annealed in a magnetic field at 200°C along the MgO[110] ∥ Fe[100] direction. EB-lithography, photolithography, and Ar-ion etching were used to pattern the stacks into 6–39 $\mu m^2$ area elliptical junctions with a long axis parallel to the Fe[100] easy axis. The MTJs were measured using a standard dc 4-probe measurement method. A physical property measurement system (Quantum Design, Dynacool) apparatus was used for temperature-dependent measurements. The TMR ratio is defined as ($R_{AP}$ − $R_P$)/$R_P$ × 100%, where $R_{P(AP)}$ is the resistance in the P (AP) magnetization state. A negative bias voltage corresponds to electrons tunneling from the bottom to the top interface. The data from Ref. [15] were used to compare the Fe/MAO/Fe to our previous Fe/MgO/Fe prepared using the same deposition systems.

The $d_{bar}$ dependence of the TMR ratio [(a) and (e)] and resistance area products ($RA$s) for the P ($RA_P$) and AP states ($RA_{AP}$) [(b) and (f)] at RT are shown in Fig. 2 for Fe/MAO/Fe and Fe/MgO/Fe, respectively (bias voltage < 10 mV). The exponential fits by $\exp(a_{P(AP)}d_{bar} + b_{P(AP)})$ for (b) and (f), where $a_{P(AP)}$ and $b_{P(AP)}$ are the slopes, and intercept for the P (AP) state are shown by blue linear lines. The fit results are summarized in Table I. The exponential background-corrected $RA_{AP}$ [$RA_P$] is shown in Figs. 2(c) and (g) [(d) and (h)]. The TMR ratios for both MTJs show clear oscillation with $d_{bar}$. The oscillation periods were almost identical (~0.31 nm), which is similar to 0.317 nm in Fe/MgO/Fe(001) by Matsumoto et al. [25] Both MTJs have similar slopes ($a_P$ ~ $a_{AP}$) for exponential changes of $RA_{AP}$ and $RA_P$ on $d_{bar}$, resulting in TMR saturation behavior at the high $d_{bar}$ region. The slopes of Fe/MAO/Fe are slightly lower than those of Fe/MgO/Fe, implying that the MAO barrier has a lower effective barrier height than the MgO barrier.[9,26] Fe/MAO/Fe shows a maximum TMR ratio (oscillation peak-to-valley difference) of 429% (125%), which is larger than the value of 417% (80%) in the Fe/MgO/Fe.[15] Therefore, MAO improved the TMR ratio and the oscillation amplitude. Notably, the TMR ratio of the Fe/MAO/Fe MTJ demonstrates sawtooth-like curves, in contrast to the sinus-like curve of the Fe/MgO/Fe MTJ. This feature can be attributed to the much larger oscillation amplitudes, sharper peak shape, and slight difference in the exponential background in the corrected $RA_P$ and $RA_{AP}$ plots [see Figs. 2 (c) and (d)]. The origin of the oscillation is still under debate;[9,27–29] nevertheless, the observed behavior indicates that the $RA$ oscillations in the P and AP states are more primitive than the TMR oscillation. Therefore, rather than the TMR ratio, background-corrected $RA$s should be used to analyze the oscillation behavior.

The temperature dependences of a Fe/MAO/Fe ($d_{bar}$ = 1.80 nm) and Fe/MgO/Fe MTJ ($d_{bar}$ = 2.18 nm) are shown in Figs. 3 (a)–(c) (bias voltage < 10 mV). (a) The TMR ratio, (b) $R_{AP}$, and (c) $R_P$. $R_{AP}$ and $R_P$ are normalized to their respective values at 300 K. At 10 K, the TMR ratio of the Fe/MAO/Fe increases up to 1,034%, which is much larger than the maximum value of 914% of the previous Fe/MgO/Fe.[15] The Fe/MAO/Fe TMR-magnetic field loops at RT and LT, show that exchange spin-valve type loops are obtained at both temperatures shown in Fig. 3 (d). By using the Jullière formula,[30] TMR = $2P_0^2/(1-P_0^2)$, the effective tunneling spin polarization $P_0$ of the Fe/MAO/Fe at LT is 0.915, which is greater than 0.905 of the Fe/MgO/Fe. This indicates that the $\Delta_1$ bands at the top and bottom Fe/MAO interfaces are approaching an ideal half-metallic state. The temperature dependence of $R_{AP}$ ($R_P$) of the Fe/MAO/Fe shows a stronger increase (weaker decrease) with temperature. The monotonical decrease in $R_P$ with reduced temperature is a common feature in high-quality Fe/MgO/Fe and Fe/MAO/Fe MTJs with a large TMR ratio >200%.[15,20] The stronger dependence in $R_{AP}$ of the Fe/MAO/Fe is because of the improved TMR ratio at LT.

The bias voltage dependences of the MTJs of Fig. 3 are shown in Fig. 4 according to the following: (a) and (d) normalized TMR ratio by its zero-bias value, (b) and (e) differential conductance $G_{AP}$, and (c) and (f) $G_P$ at RT (300 K) and LT (10 K for Fe/MAO/Fe and 5 K for Fe/MgO/Fe), respectively. The curves of the Fe/MAO/Fe are more symmetric with the bias polarity than those of the Fe/MgO/Fe, indicating that the top and bottom Fe/MAO interfaces are almost identical in their electronic states. Therefore, our process optimization effectively controls the degree of oxidation and crystallinity of the MAO interfaces. The bias voltage dependence of the TMR ratio is similar for both MTJs; thus, the voltage at which the TMR ratio reduces to 50% of its zero-bias value, $V_{half}$, is also similar (~0.8 V at RT). The $G_{AP}$ spectra of both MTJs show a similar parabolic shape. The Fe/MAO/Fe shows a bit steeper parabolic shape than the Fe/MgO/Fe. The magnon excitation causes a large zero-bias dip at LT, resulting in steep slopes near zero-bias in the normalized TMR curves seen in various MTJs.[31–33] Broad minima are observed in the $G_P$ spectra at RT for positive and negative bias voltage. At LT, the minima continue to develop at both bias polarities, as indicated by the arrows in Fig. 4 (f). Fine peaks around |0.2| and |0.4| V of the Fe/MAO/Fe are clearer than those in the Fe/MgO/Fe. In addition to the fine structures, additional shoulders appear at higher bias voltage around |0.8–1.0| V in the Fe/MAO/Fe. These fine structures originate from the formation of distinct interfacial electronic states at the barrier/Fe interfaces, indicating improved barrier interface crystallinity and flatness in the Fe/MAO/Fe. This result is consistent with the observed increases in maximum TMR ratio and TMR oscillation amplitude.

The used MAO can behave as MgO doped with a trace of Al, resulting in a slight reduction of the lattice spacing of the MgO barrier. The lattice constant $a$ of $Mg_4Al\text{-}O_x$ is estimated to be ~ 0.416 nm assuming Vegard's law of MgO ($a$ = 0.421 nm) and cation-disordered $MgAl_2O_4$ (=half of a $MgAl_2O_4$ lattice ~ 0.404 nm), assuming that the barrier composition is unchanged from the evaporation source.[19] Therefore, the lattice mismatch with Fe ($\sqrt{2}a$ = 0.405 nm for 45° in-plane rotation) is expected to be 2.6%, smaller than 3.8% with MgO. Although nano-structural analysis must evaluate an actual mismatch in our Fe/MAO/Fe MTJs, the reduced dislocation density using MAO is expected to be the origin of the improved TMR properties and clearer fine peaks in the $G_P$ spectrum at LT.

However, we did not observe any significant improvement of $V_{\text{half}}$, unlike the previously reported MAO-based MTJs.[17,20,22] This indicates that tuning the Al amount of an MAO barrier and the interface modification process can further improve the Fe/MAO/Fe(001) structure.

In summary, we demonstrated large RT and LT TMR ratios of up to 429% and 1,034%, respectively, in an MTJ with a Fe/Mg$_4$Al-O$_x$/Fe(001) structure. Additionally, significantly large TMR oscillation was also observed as a function of barrier thickness, with an oscillation peak-to-valley difference reaching 125% at RT. The observed TMR ratios and TMR oscillation amplitude exceeded the values of the previous Fe/MgO/Fe(001) MTJs. The features of the differential conductance spectra, such as deep minima and well-developed fine structures, indicate improved barrier interface states by the lattice constant tuning of the MAO barrier. This study shows that much larger TMR ratios can be expected with the development of barrier materials and improved interfaces, which may lead to novel spintronic applications.

**Acknowledgements**

The authors are grateful to H. Ikeda for her technical support on device microfabrication. The work was partly supported by the JSPS KAKENHI Grant Nos. 21H01750 and 21H01397. This paper is partly based on results obtained from a project, JPNP16007, commissioned by the New Energy and Industrial Technology Development Organization (NEDO).

**Table I.** Fitting parameters of the exponential background $RA_{P(AP)} = \exp(a_{P(AP)}d_{bar} + b_{P(AP)})$ in Figs. 2 (b) and (f).

|      | $a_{AP}$ (nm$^{-1}$) | $b_{AP}$ | $a_P$ (nm$^{-1}$) | $b_P$ |
|------|----------------------|----------|-------------------|-------|
| MgO  | 6.19                 | −1.95    | 6.01              | −3.16 |
| MAO  | 5.68                 | −1.11    | 5.51              | −2.28 |

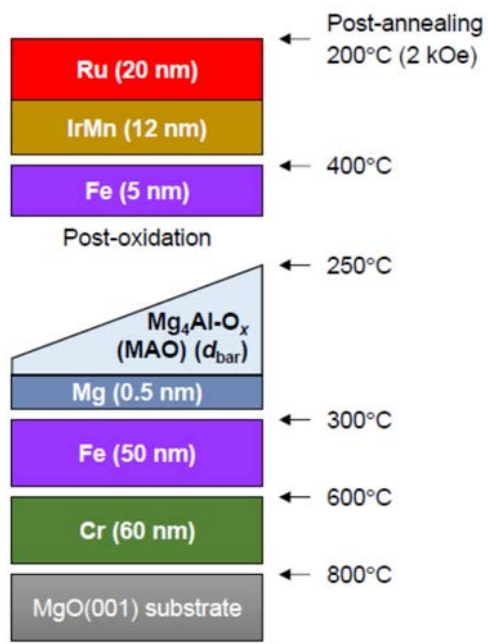

**FIG. 1.** Schematic Fe/MAO/Fe(001) MTJ stacking structure with post-annealing temperatures.

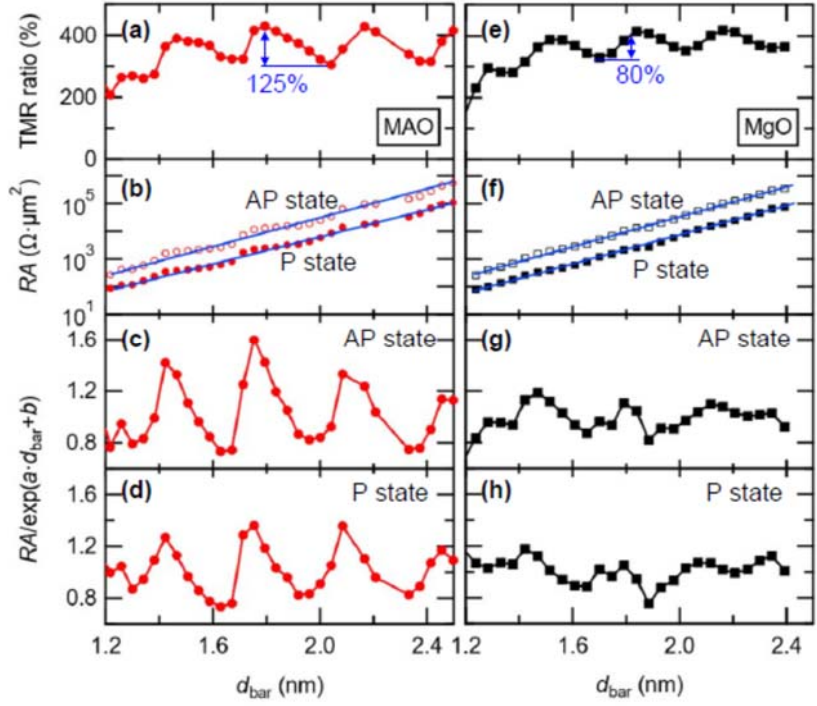

**FIG. 2.** $d_{Bar}$ dependences of TMR properties for (a)-(d) Fe/MAO/Fe and (e)-(h) Fe/MgO/Fe (Ref. [15]) MTJs at RT. (a), (e) TMR ratio and (b), (f) $RA_{AP}$ and $RA_P$. (c), (g) [(d), (h)] Exponential background corrected $RA_{AP}$ [$RA_P$]. Blue lines in (b) and (f) are exponential fits of $RA_{AP}$ and $RA_P$. The Fe/MgO/Fe data were reproduced from T. Scheike *et al.*, Appl. Phys. Lett. **118**, 042411 (2021), with the permission of AIP Publishing.

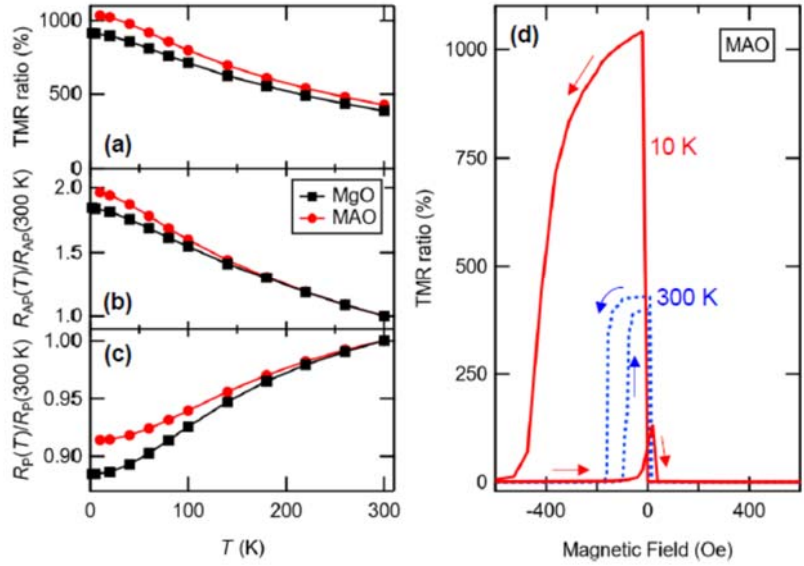

**FIG. 3.** (a)-(c) Temperature dependences of (a) TMR ratio, (b) $R_{AP}$, and (c) $R_P$ of Fe/MAO/Fe (circle) and Fe/MgO/Fe (square, Ref. [15], reproduced from T. Scheike *et al.*, Appl. Phys. Lett. **118**, 042411 (2021), with the permission of AIP Publishing). For (b) and (c), $R_{AP}$ and $R_P$ were normalized by their 300 K values. (d) TMR-magnetic field loops at 300 K and 10 K of Fe/MAO/Fe.

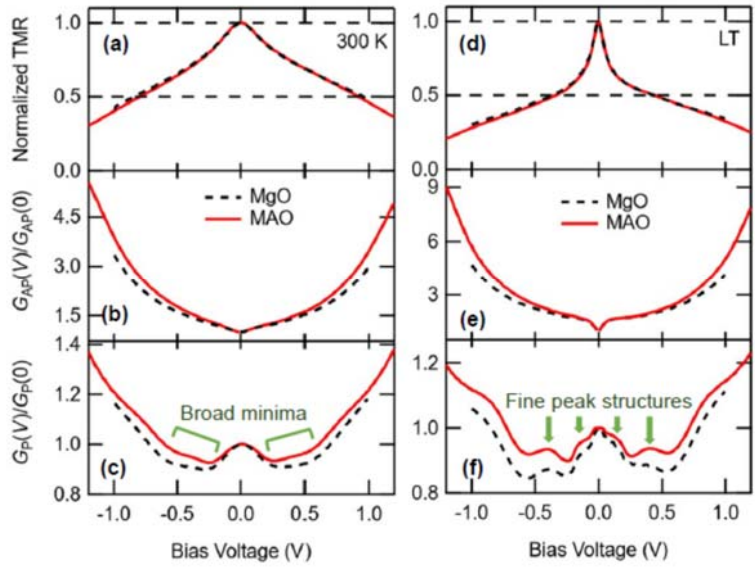

**FIG. 4.** Bias voltage dependences of (a) [(d)] normalized TMR ratio, normalized conductance (b) [(e)] $G_{AP}$, and (c) [(f)] $G_P$ at 300 K [LT] of Fe/MAO/Fe (red solid) and Fe/MgO/Fe (black dashed) (Ref. [15], reproduced from T. Scheike *et al.*, Appl. Phys. Lett. **118**, 042411 (2021), with the permission of AIP Publishing).